\documentclass{iopart}

\usepackage{psfig}

\font\teuf=eufm10 scaled 1200
\font\seuf=eufm7 scaled 1200            
\font\sseuf=eufm5 scaled 1200

\newfam\euffam
        \textfont\euffam=\teuf
        \scriptfont\euffam=\seuf
        \scriptscriptfont\euffam=\sseuf

\newtheorem{theo}{Theorem}

\def\S0{ \{ S_0 \} }

\def\AC{{\mathcal A}}

\def\MC{{\mathcal M}}

\def\PC{{\mathcal P}}
\def\QC{{\mathcal Q}}

\def\vect#1{\mbox{\boldmath{$#1$}}}

\def\ra{\rightarrow}

\def\maxindex#1{{#1}_{\mbox{\scriptsize max}}}
\def\minindex#1{{#1}_{\mbox{\scriptsize min}}}

\font\tenbb=bbold10 scaled \magstep1

\font\sevenbb=bbold7 scaled \magstep1

\font\fivebb=bbold5 scaled \magstep1

\newfam\bbfam
\textfont\bbfam=\tenbb
\scriptfont\bbfam=\sevenbb
\scriptscriptfont\bbfam=\fivebb
\def\bb{\fam\bbfam\tenbb}

\def\Zb{\mbox{$\bb Z$}}

\def\1b{\mbox{$\bb 1$}}

\def\openone{{\leavevmode\hbox{\small1\kern-3.55pt\normalsize1}}}
\def\tP{\tilde{P}}
\def\eeq#1{\vect{e}^{\mbox{\scriptsize eq}}_{#1}}

\def\scal#1#2{\langle {#1} | {#2} \rangle}
\newtheorem{conj}{Conjecture}
\begin{document}

\noindent {\sc Preprint} \hfill  \today

\title[\bf Spectral gaps of Markov chains: a multi-decomposition
technique]{\bf Bounding spectral gaps of Markov chains: a novel
exact multi-decomposition technique}

\author{{\sc N. Destainville} \medskip \\
Laboratoire de Physique Quantique~-- UMR CNRS/UPS 5626 \medskip \\
Universit\'e Paul Sabatier, \medskip \\ 
118, route de Narbonne, 31062 Toulouse Cedex 04, France.}

\begin{abstract} We propose an exact technique to calculate 
lower bounds of spectral gaps of discrete time reversible Markov
chains on finite state sets. Spectral gaps are a common tool for
evaluating convergence rates of Markov chains. As an illustration, we
successfully use this technique to evaluate the ``absorption time'' of
the ``Backgammon model'', a paradigmatic model for glassy dynamics.
We also discuss the application of this technique to the ``Contingency
table problem'', a notoriously difficult problem from probability
theory. The interest of this technique is that it connects spectral
gaps, which are quantities related to dynamics, with static
quantities, calculated at equilibrium.
\end{abstract}

{\small

}

\bigskip

Markov chains~\cite{Grimmett,Feller,Sinclair93,Jerrum96} have
applications in many areas, ranging from pure probability theory to
theoretical or numerical statistical physics. For example, in
out-of-equilibrium statistical physics, they are commonly used to
write in a formalized form the time evolution of physical
models~\cite{Newman}. We study in this paper~-- as a particular
application of our technique~-- the ``Backgammon
model''~\cite{Ritort95} which is a well-known paradigmatic mean-field
model for glassy dynamics. Markov chains are also wide-spread in
numerical statistical physics: Monte Carlo algorithms~\cite{Newman}
with Metropolis or Glauber dynamics, which are examples of Markov
processes, are of great importance in the computational study of
complex systems. The efficiency of such algorithms relies on their
rates of convergence towards their equilibrium distributions. The
ergodic times must be as short as possible to save computational
time. Usually, such algorithms are said to be {\em rapid} if their
ergodic times are polynomial in the system sizes whereas (generally
speaking) the numbers of configurations grow exponentially with system
sizes. Monte Carlo algorithms are also more and more wide-spread in
applied mathematics~(see \cite{Jerrum96}): the ``Contingency table
problem''~\cite{Diaconis85,Barlow} that we examine at the end of the
present paper will provide an example.

Roughly speaking a Markov $\MC$ process needs a characteristic time
$\tau$ to be close to equilibrium (the equilibriation time).  If $P$
denotes the transition matrix (defined below) associated with $\MC$,
if $1/g(P)$ is the spectral gap of $P$, that is to say the difference
between the two largest eigenvalues of $P$, then $\tau \sim
1/g(P)$. We explain this point and we give references below.
Therefore if one calculates a lower bound on $g(P)$, one also gets an
upper bound on $\tau$. More precisely, if $1/g(P)$ is polynomial in
the system size, then $\tau$ is also polynomial and the chain is
rapid.

Many efficient techniques have been developed to estimate (or to
bound) mixing times and spectral gaps, when elementary techniques from
linear algebra do not apply. Among many others, we mention the
``Coupling'' technique (see \cite{Aldous81}), the ``Canonical path''
and ``Conductance'' arguments (see~\cite{Sinclair93,Jerrum96}), or the
``Chain decomposition'' method~\cite{Randall}.  This last method
relies on the following basic idea: (i) decompose the state space into
disjoint smaller pieces; (ii) prove that the dynamics is rapid on each
piece considered in isolation; (iii) re-compose the dynamics on the
whole state space from that on pieces and prove it is rapid in its
turn.

We insist on this method because our ``Multi-decomposition'' technique
is an extension of the previous one. More precisely, the idea of the
chain decomposition method is schematically the following: suppose
that the state space $\Omega$ of the Markov chain can be ``naturally''
decomposed into several disjoint subsets $\Omega_a$, $a=1,\ldots,A$;
then ``cut'' all the transitions between different subsets $\Omega_a$
and only keep the transitions inside each subsets. The so-obtained
restricted Markov chain on each $\Omega_a$ is denoted by
$\MC_a$. Suppose then that one can prove that each $\MC_a$ is rapid on
$\Omega_a$, and that in addition the original dynamics is rapid
``between'' subsets. Then, under certain conditions, the original
dynamics on $\Omega$ will also be rapid.  Unfortunately, even if one
can easily prove that the dynamics is rapid on each $\Omega_a$, it can
be extremely difficult or even impossible to handle the second step of
the proof. In reference~\cite{PRLbibi}, an idea which bypasses this
difficulty was presented in the case of random rhombus tilings. In the
present paper, we generalize this idea in a formalized form and we
show that it has a much vaster domain of application than random
tilings.

\begin{figure}[ht]
\begin{center}
\ \psfig{figure=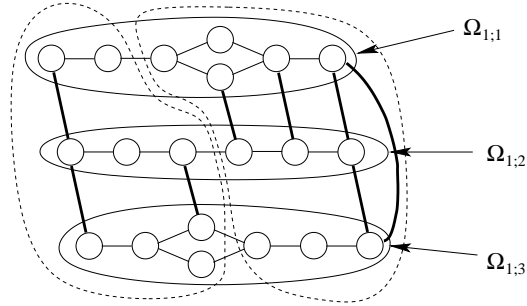,height=4cm} \
\end{center}
\caption{Schematic representation of a multi-decomposition 
in the case of 2 decompositions. There are 20 possible states in
$\Omega$ represented by small circles, and the edges between these
circles represent possible transitions between states. The first
decomposition contains 3 subsets $\Omega_{1;a}$ (full lines) and the
second one, 2 subsets $\Omega_{2;b}$ (dashed lines).  In the first
decomposition, we have emphasized (thick lines) the edges that are cut
because the transitions between states of different subsets
$\Omega_{1;a}$ are forbidden. If one can prove that the dynamics is
rapid in each subset of each decomposition then, under certain
conditions, one can conclude that the dynamics is rapid on the whole
state set $\Omega$.}
\label{decomps}
\end{figure}

As compared to the previous ``simple'' decomposition method, the
present ``multi-decomposition'' scheme uses several ($m$)
decompositions of $\Omega$, namely
$(\Omega_{1;a_1}),(\Omega_{2;a_2}),\ldots,(\Omega_{m;a_m})$.  For each
decomposition indexed by $k=1,\ldots,m$, the set $\Omega$ is also
decomposed in disjoint subsets $\Omega_{k;a}$. As in the
decomposition method, one first needs to prove that the dynamics is
rapid in each subset $\Omega_{k;a}$, for each $k$ and each $a$.  Then
if any two decompositions ``overlap sufficiently'', in a sense that
will be precised below, the dynamics will also be rapid on the whole
set $\Omega$. In other words, given a decomposition $(\Omega_{k;a})$,
one does not have to prove any more that the original dynamics is
rapid between subsets of this decomposition. This point is ensured
instead by the remaining subsets $\Omega_{l;b}$, $l \neq k$. These
ideas are schematically exemplified in figure~\ref{decomps}.

\begin{figure}[ht]
\begin{center}
\ \psfig{figure=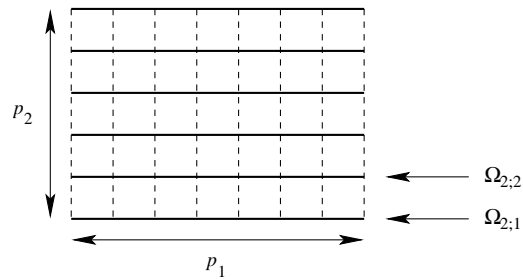,height=3.6cm} \
\end{center}
\caption{A 2D random walker on a square grid $p_1 \times p_2$. At time
$t$, the walker is situated on a vertex. At time $t+1$, he chooses
with the same probability to move on a nearest neighbor, horizontally
(full lines) or vertically (dashed lines), if he can. There are two
natural decompositions of this Markov chain, a vertical and a
horizontal one. In the horizontal decomposition, for example, the
walker can only move horizontally on subsets $\Omega_{2;b}$. Vertical
edges are ``cut''. }
\label{rw2D}
\end{figure}

We illustrate this delicate point with a simple example that will also
be developed in greater detail below: consider a two-dimensional
random walker on a square grid $p_1 \times p_2$ (see
figure~\ref{rw2D}). At each step, the walker performs either a
vertical or a horizontal move, with equal probability 1/4 in each
direction (we neglect the question of boundaries at this level of 
discussion). Then in our formalism, we define two decompositions
$(\Omega_{1;a})$ and $(\Omega_{2;b})$, respectively a vertical and a
horizontal one. In the vertical decomposition, each subset
$\Omega_{1;a}$, $a=1,\ldots,p_1$, is a vertical segment, with no
possible transitions towards the neighboring vertical segments: the
horizontal edges are ``cut'' in this vertical
decomposition. Symmetrically, vertical edges are ``cut'' in the
horizontal decomposition, and the subsets $\Omega_{2;b}$,
$b=1,\ldots,p_2$ are horizontal segments.  Now on each subset
$\Omega_{1;a}$ or $\Omega_{2;b}$, the induced dynamics on vertical or
horizontal segments is nothing but that of a one-dimensional random
walker which stays at the same place with probability 1/2 and moves in
either direction with probability 1/4. For such a random walker on
segments of lengths $p_k$, the spectral gaps are $g_k \simeq Cst/p_k^2$: the
dynamics is rapid on each subset of each decomposition. Then the
method developed in this paper can be applied to this problem. The
underlying idea is that any trajectory of the walkers is a succession
of vertical and horizontal moves. Therefore the dynamics on $\Omega$
is a combination of the dynamics on the different subsets. 
If the dynamics is rapid on each subset, it will certainly be rapid on
$\Omega$. More formally, one proves {\em via} the present technique that
\begin{equation}
g(P) \geq \inf(g_1,g_2) \simeq \frac{Cst}{\sup(p_1^2,p_2^2)}.
\end{equation}
In other words, the dynamics on the whole set $\Omega$ is faster than
the dynamics on the slowest subset. In this case, this inequality can
be checked by independent means and appears to be an equality. In the
present paper, we rigorously formalize these ideas in a more general
point of view.

\medskip

The interest of this method is that it can be applied
inductively. This idea will be used throughout the paper. In the
previous example, one can iterate the multi-decomposition scheme in
the case of random walkers on larger and larger $m$-dimensional
hyper-cubic grids of side lengths $p_1 \times \ldots \times p_m$.  At
each stage $m$, there are $m$ decompositions, one for each dimension
of space, and the subsets are $(m-1)$-dimensional hyper-cubic grids in
their turn (see figure~\ref{walker3D}). Then one proves by induction
on $m$ that
\begin{equation}
g(P^{(m)}) \geq \inf_{k=1,\ldots,m} g(P^{(m-1)}_k)
\simeq \frac{Cst}{\sup_k(p_k^2)},
\end{equation}
where $P^{(q)}$ is the transition matrix of a $q$-dimensional random
walker. Once again, this lower bound is in fact exact.

The organization of this paper is as follows: the first section
introduces the background definitions and notations used throughout the
paper. In particular, the notion of discrete time reversible Markov
chain on a finite state set is briefly recalled. The second section
presents the specific definitions and the main results of the paper,
which are applied to several pedagogical elementary examples in section 3,
namely random walkers on hyper-cubic lattices and on the
symmetric group $S_n$.

In section~\ref{urn}, we successfully apply the present technique to
two ``Urn models'', the ``Backgammon model'' and the ``Monkey urn
model''. In both cases, we relate the dynamics on $m$ urns to the
dynamics on 2 urns. We show that the source of the slow dynamics of
the Backgammon model is entirely contained in the 2-urn problem. In
other words, considering many urns instead of 2 does not bring any
additional slowness~\cite{Lipowski97,Murthy97}. At last, the
Contingency table problem is analyzed in section~\ref{CTP}. The last
section~\ref{discussion} is devoted to conclusions.

\section{Generalities}
\label{generalities}

\subsection{Transition matrix $P$ of a Markov chain $\MC$}

The present paper deals with discrete time reversible Markov chains on
finite state sets. Let $\Omega$ be a finite set of cardinality $N$,
$\MC$ a Markov process on $\Omega$, and $P$ its transition matrix of
transition probabilities $P(x,y)$: $P(x,y) \geq 0$ is the conditional
probability that the chain is in the state $x$ at time $t+1$ given that
it was in the state $y$ at time $t$. Note that these transition
probabilities themselves do not depend on time. Probabilities are
conserved so that $\sum_x P(x,y)=1$ for all $y$: $P$ is {\em
stochastic}.

We define the state vector $\vect{e}(t)$ at time $t$: it has coordinates
$(e(t,x))_{x \in \Omega}$ where $e(t,x)$ is the probability that
the chain is in the state $x$ at time $t$. Then by definition of
the matrix $P$, 
\begin{equation}
\vect{e}(t+1) = P \vect{e}(t).
\end{equation}
A state vector $\vect{\pi}$ is said to be an equilibrium (or
stationary) distribution if $\vect{\pi} = P \vect{\pi}$, in other
words if \vect{\pi} is an eigenstate of $P$ with eigenvalue 1.

The Markov process $\MC$ is said to be reversible if it satisfies the
detailed balance condition~\cite{Grimmett,Newman}
\begin{equation}
\pi(x) P(y,x) = \pi(y) P(x,y)
\label{bilan}
\end{equation}
for all states $x$ and $y$. This condition ensures that $\vect{\pi}$ is
effectively an equilibrium distribution for $\MC$. It will be useful
in the following. Furthermore, we suppose that this equilibrium
distribution $\vect{\pi}$ exists and is unique (see for instance
\cite{Grimmett} for further detail on this point). In addition, we
assume in the sequel that the ``loop'' transitions $P(x,x)$ are at
least $1/2$ for all states $x$. 

\subsection{Eigenvalues and spectral gap of $P$}

We basically assume for the moment that $\pi(x) > 0$ for all states
$x$: the chain can reach any state at equilibrium. Then we define the
scalar product:
\begin{equation}
\scal{\vect{f}}{\vect{g}} \equiv \sum_{x \in \Omega}
\frac{1}{\pi(x)} f(x) g(x).
\label{scalprod}
\end{equation}	
Because of reversibility, $P$ is self-adjoint for this scalar product:
\begin{eqnarray}
\scal{P\vect{f}}{\vect{g}} & = & \sum_x \frac{1}{\pi(x)} \sum_y 
P(x,y) f(y) g(x) \\
& = & \sum_y \frac{1}{\pi(y)} \sum_x 
P(y,x) g(x) f(y) \qquad \mbox{(see eq. (\ref{bilan}))}\\
& = & \scal{\vect{f}}{P\vect{g}}.
\end{eqnarray}
Thus the eigenvalues $\beta_i$, $i=0,\ldots,N-1$, of $P$ are
real. Moreover, $|\beta_i|\leq 1$ for all $i$ by standard
Perron-Frobenius theory~\cite{Mehta}. Since the equilibrium
distribution exists and is unique, $\beta_0=1$ and
\begin{equation}
1 > \beta_1 \geq \beta_2 \geq \ldots \geq \beta_{N-1} \geq -1.
\label{ordrevap}
\end{equation}
In addition, we have supposed that $P(x,x) \geq 1/2$ for all $x$. As a
consequence, $\tilde{P}=2 P - \mbox{\openone}$ is also a transition
matrix with non-negative transition probabilities, the eigenvalues $2
\beta_i -1$ of which are larger than $-1$ according to
(\ref{ordrevap}).  Hence $\beta_i \geq 0$ for all $i$. Now we can
define the {\em spectral gap} of $P$:
\begin{equation}
g(P) = 1 - \beta_1 >0.
\end{equation}

\subsection{Spectral gap and equilibriation time $\tau$}

In the introduction, we claimed that the characteristic time $\tau$ to
be close to equilibrium is of order $1/g(P)$, whatever the initial
probability distribution $\vect{e}(0)$. Let us precise this point and
in particular what we mean by ``close''. If the chain was in the state
$x\in \Omega$ at time 0, we denote by $\vect{e}_x(t)$ the state vector 
at time $t$.

In the literature, the distance between the probability distribution
$\vect{e}_x(t)$ and the equilibrium one $\vect{\pi}$ can be measured
by various means. Among them, two norms are commonly used, the
Euclidean norm associated with the above scalar product and the 1-norm
or {\em variation distance}~\cite{Aldous81}:
\begin{equation}
\Delta_x(t) = \frac{1}{2} \sum_{y \in \Omega} | e_x(t,y) -\pi(y) |.
\end{equation}
Note that $\Delta_x(t) = \sup_{A \in \Omega} |p(A,t) - \pi(A)|$
where $p(A,t)$ (resp. $\pi(A)$) is the probability that the system
is in the subset $A$ of $\Omega$ at time $t$ (resp. at equilibrium).
Then given $\varepsilon >0$, we define the {\em mixing} time
$\tau_{var} (\varepsilon)$ associated with this distance:
\begin{equation}
\tau_{var} (\varepsilon) =\max_{x} \min_{t_0} \left\{ t_0 / \forall
t \geq t_0, \Delta_x(t) \leq \varepsilon \right\}.
\end{equation}
In other words, whatever the initial state $x$, one is sure to stay
within distance $\varepsilon$ of equilibrium after $\tau_{var}
(\varepsilon)$ steps. Then elementary algebra (see
e.g.~\cite{Diaconis91}) shows that:
\begin{equation}
\tau_{var} (\varepsilon) \leq \left( \ln \left(\frac{1}{ \varepsilon}
\right) + \max_x \ln \left(\frac{1}{\pi(x)} \right) \right) \
\frac{1}{g(P)}.
\label{epsvar}
\end{equation}
This upper bound is of order $1/g(P)$ even though the second term of
the prefactor of $1/g(P)$ can be substantially big. However, this
second term is usually polynomial in the system size, at least at high
temperature, where the distribution $\vect{\pi}$ is uniform: this
second term is simply the entropy of the system and if it is
extensive, it is polynomial in the system size. If $1/g(P)$
is also polynomial, the dynamics remains rapid. The reader can refer
to Wilson~\cite{Wilson01} for an interesting discussion on this point.

On the contrary, at vanishing temperature, only the low energy levels
are occupied and some probabilities $\pi(x)$ tend to zero and this
second term diverges, even for finite systems. This is not physically
relevant, since $\tau_{var}$ remains finite in general. The previous
upper bound (\ref{epsvar}) is not adapted.

Should we have chosen the Euclidean norm to measure the distance under
interest, we would have been led to the similar conclusion that the
equilibriation time is of order $1/g(P)$ (see also \cite{Wilson01} for
further discussion). In the sequel of this paper, we shall focus on
spectral gaps rather than such equilibriation times.

\section{Main result}
\label{main.result}

\subsection{Definitions}
\label{defs}

Generally speaking, given a discrete time Markov chain $\MC$ on a
finite set $\Omega$, a decomposition~\cite{Randall} of $\MC$ is both a
partition of $\Omega$ in disjoint subsets $\Omega_a$, $a=1,\ldots,A$,
and a new natural dynamical rule on $\Omega$. This dynamical rule is
defined by its transition matrix $P'$ as follows: for all $x
\neq y \in \Omega$, if $x$ and $y$ belong to the same subset
$\Omega_a$ then $P'(x,y) = P(x,y)$, else $P'(x,y)=0$. We say that we
``cut transitions between the different subsets $\Omega_a$''. The diagonal
terms $P'(x,x)$ are suitably modified to ensure that $P'$ remains
stochastic. If $M_a$ denotes the restriction of $P'$ to $\Omega_a$,
then $P'$ is block-diagonal and it is the direct sum of the $M_a$:
\begin{equation}
P'=M_1 \oplus M_2 \oplus \ldots \oplus M_A. 
\label{sommedir}
\end{equation}
We naturally suppose that the decomposition is designed so that each
subset $\Omega_a$ is connected with respect to the new dynamical rule
and that for each $\Omega_a$ seen in isolation, there also exists a
unique stationary distribution. Then the equilibrium state of $P'$ is
degenerate of dimension $A$. We call $P'$ the {\em restriction} of $P$
to the decomposition $(\Omega_a)$.

As it was announced in introduction, we shall consider the case where
we have $m$ different decompositions
$(\Omega_{1;a_1}),(\Omega_{2;a_2}),\ldots,(\Omega_{m;a_m})$,
$a_k=1,\ldots,A_k$. {\em A priori}, the different $A_k$ need not to be
equal. Then we define as above $m$ different restrictions for each
decomposition $(\Omega_{k;a})$, denoted by $P_k$. We say that we
have a multi-decomposition or a $m$-decomposition.

Now we need to introduce a property of regularity.  We say that a
transition of the original chain $P(x,y) > 0$ ``belongs'' to the
decomposition $(\Omega_{k;a})$ if $P_k(x,y) > 0$, in other words if
this transition is not cut in that decomposition.

\medskip

\noindent {\bf Definition} {\em A $m$-decomposition
$(\Omega_{1;a_1}),(\Omega_{2;a_2}),\ldots,(\Omega_{m;a_m})$ of $\MC$
is said to be {\em regular} if there exists an integer $r\leq m$ such
that each transition $P(x,y) > 0$ belongs to exactly $r$
decompositions. The integer $r$ is called the {\em redundancy} of the
regular multi-decomposition.}

\medskip

If the $m$-decomposition is regular of redundancy $r$ then we have
the following simple but important relation:
\begin{equation}
P = \frac{P_1 + P_2 + \ldots + P_m}{r} + \frac{r-m}{r} \; \openone,
\end{equation}
which interconnects the different decompositions~\cite{PRLbibi}.  The
second term of the sum restores the diagonal coefficients $P(x,x)$.

Note that the restriction of $\MC$ to each subset $\Omega_{k;a}$ of
each decomposition remains reversible, because if $P(x,y)$ is cut,
then $P(x,y)$ is also cut:
\begin{equation}
\pi(x) P_k(y,x) = \pi(y) P_k(x,y)
\label{balance.k}
\end{equation}
In addition, $\vect{\pi}$ remains a stationary distribution for each
$P_k$, but it is not unique; We denote by $E_k$ the eigenspace of
eigenvectors of $P_k$ corresponding to the eigenvalue 1. Its dimension is
$A_k$ because we also suppose that on each subset $\Omega_{k;a}$ there
is a unique stationary distribution. Equation~(\ref{balance.k})
also leads us to the conclusion that $P_k$ remains self-adjoint for
the scalar product (\ref{scalprod}). If we denote by $\Pi_k$ the
orthogonal projection (with respect to this scalar product) onto
$E_k$, then any eigenvector (other than vectors of $E_k$) projects onto
$\vect{0}$. We define the {\em multi-projection}:
\begin{equation}
\Pi = \Pi_1 + \Pi_2 + \ldots + \Pi_m.
\end{equation}
Then $\Pi (\vect{\pi}) = m \vect{\pi}$. We say that the
multi-decomposition is {\em non-degenerate} if this eigenvector is
non-degenerate himself, which is satisfied in practice. We denote by
$\nu$ the second largest eigenvalue of $\Pi$ (in modulus) and we call
it the {\em norm of the multi-projection $\Pi$}. In the following, the
difficult step will often be to calculate this norm. We shall discuss
this point in subsection~\ref{Howto} and give several examples in
sections~\ref{e.examples}, \ref{urn} and \ref{CTP}.

\medskip

\noindent {\bf Definition} {\em A non-degenerate $m$-decomposition,
which is regular of redundancy $r$ and which has an associated 
multi-projection of norm $\nu$, is called a non-degenerate
$(m,r,\nu)$-multi-decomposition.}

\subsection{Main theorem}
\label{main.theo}

The eigenspace $E_k$ of $P_k$ corresponding to the eigenvalue $1$ is
degenerate: we denote by $g(P_k)$ the difference between 1 and the 
second largest eigenvalue of $P_k$ (strictly smaller than 1).

\medskip

\begin{theo}
\label{th1}
Let $P$ be the transition matrix of a Markov chain $\MC$ on $\Omega$
and $(\Omega_{1;a_1}),(\Omega_{2;a_2}),\ldots, (\Omega_{m;a_m})$ be a
non-degenerate $(m,r,\nu)$-multi-decomposition of $\MC$. If $P_k$
denotes the restriction of $P$ to the decomposition
$(\Omega_{k;a_k})$, then the following {\em gap relation} holds:
\begin{equation}
g(P) \geq \frac{m-\nu}{r} \inf_k g(P_k).
\label{gaprel}
\end{equation}
\end{theo}

\medskip

\noindent {\em Proof:} For the sake of convenience, we work in this
proof with the matrix:

\begin{equation}
\tilde{P} = {P_1 + P_2 + \ldots + P_m \over m} = \frac{r}{m} P 
+ \frac{m-r}{m} \openone. 
\end{equation}
At the end of the calculation, the gap of $P$
will simply be: $g(P)=m/r \ g(\tilde{P})$.

Given a state vector $\vect{e}$, we note $\eeq{k}=\Pi_k(\vect{e})$
(the superscript ``eq'' stands for ``equilibrium''); $\eeq{k}$ depends
on $k$ and $\vect{e}$, since the eigenspace of $P_k$ corresponding to
the eigenvalue 1 is degenerate. All the difficulty in the following
lies in this degeneracy.

We suppose now that we have sorted altogether all the eigenvalues of
the $m$ matrices $P_k$: $1 > \mu_1 \geq \mu_2 \geq \ldots
\geq \mu_q \geq 0$. We denote by $\vect{f}_j$ the normalized eigenstate
corresponding to the eigenvalue $\mu_j$; $\vect{f}_j$ can {\em a priori} be
the eigenstate of {\em any} matrix $P_k$. The vector $\vect{e} - \vect{\pi}$
is orthogonal to $\vect{\pi}$ and we write
\begin{eqnarray}
\label{zorglub}
\vect{e} - \vect{\pi} & = & {1 \over m} \sum_{k=1}^m (\vect{e} - \eeq{k}) + 
{1 \over m} \sum_{k=1}^m (\eeq{k} - \vect{\pi}) \\ \nonumber
 & = & {1 \over m} \sum_{j=1}^q \alpha_j \vect{f}_j +
{1 \over m} \sum_{k=1}^m (\eeq{k} - \vect{\pi})
\end{eqnarray}
where each vector $(\vect{e} - \eeq{k})$ has been projected
on the eigenbasis of $P_k$, and
\begin{eqnarray}
\tP(\vect{e} - \vect{\pi}) & = & {1 \over m} \sum_{k=1}^m P_k (\vect{e} -
 \vect{\pi}) \\ \nonumber
& = & {1 \over m} \sum_{k=1}^m P_k(\vect{e} - \eeq{k}) + {1
 \over m} \sum_{k=1}^m (\eeq{k} - \vect{\pi}) \\ \nonumber
& = & {1 \over m}
 \sum_{j=1}^q \mu_j \alpha_j \vect{f}_j + {1 \over m} \sum_{k=1}^m (\eeq{k} -
 \vect{\pi}),
\end{eqnarray}
by definition of $\vect{f}_j$ and $\mu_j$. Now, thanks to a suitable Abel
transform,
\begin{eqnarray}
\tP(\vect{e} - \vect{\pi}) & = & {1 \over m} (1-\mu_1)\sum_{k=1}^m 
(\eeq{k} - \vect{\pi}) \\ \nonumber
 & + & {1 \over m} \sum_{r=1}^{q-1} (\mu_r - \mu_{r+1})
 \left(\sum_{k=1}^m (\eeq{k} - \vect{\pi}) +
\sum_{j=1}^r \alpha_j \vect{f}_j \right) \\ \nonumber
 & + & {1 \over m} \ \mu_q \left(\sum_{k=1}^m (\eeq{k} - \vect{\pi}) + 
\sum_{j=1}^q \alpha_j \vect{f}_j \right).
\end{eqnarray}
Thus, if $\| \cdot \|$ stands for the Euclidean norm associated with
the scalar product~(\ref{scalprod}),
\begin{eqnarray}
\|  \tP(\vect{e} - \vect{\pi}) \| &  \leq & {1 \over m} (1-\mu_1) \|
		\sum_{k=1}^m (\eeq{k} - \vect{\pi}) \| \\ \nonumber
 &  & + \sum_{r=1}^{q-1} (\mu_r - \mu_{r+1}) \| \vect{e} - \vect{\pi} \|
+ \mu_q  \| \vect{e} - \vect{\pi} \|. \\ \nonumber
& \leq & {1 \over m} (1-\mu_1) \| 
\sum_{k=1}^m (\eeq{k} - \vect{\pi}) \|  + \mu_1 \| \vect{e} - \vect{\pi} \|.
\end{eqnarray}
Indeed, $\mu_k - \mu_{k+1} \geq 0$ and $\mu_q \geq 0$; moreover, for
any $r \leq q$, $\displaystyle{\sum_{k=1}^m (\eeq{k} - \vect{\pi}) +
\sum_{j=1}^r \alpha_j \vect{f}_j}$ is the sum of $m$ orthogonal projections of
$\vect{e} - \vect{\pi}$ on suitable spaces, the norm of each of them being
therefore smaller than $\| \vect{e} - \vect{\pi} \|$. In addition,
$\sum_k (\eeq{k} - \vect{\pi}) = \Pi (\vect{e}- \vect{\pi})$ and
by definition of $\nu$,
\begin{equation}
\| \sum_{k=1}^m (\eeq{k} - \vect{\pi}) \| \leq 
\nu \| e - \vect{\pi} \|.
\label{majo}
\end{equation}
As a consequence,
\begin{equation}
\|  \tP(\vect{e} - \vect{\pi}) \| \leq [{\nu \over m} (1 - \mu_1) + \mu_1]
\ \| \vect{e} - \vect{\pi} \|,
\end{equation}
from which it follows that
\begin{equation}
g(\tP) \geq {m - \nu \over m} (1 - \mu_1) \geq {m - \nu \over m} 
\inf_k g(P_k).
\label{genrel}
\end{equation}
Thus 
\begin{equation}
g(P) \geq {m - \nu \over r} \inf_k g(P_k).
\label{mino}
\end{equation} 

\subsection{How to calculate the norm $\nu$ of the multi-projection}
\label{Howto}

As mentioned above, the difficult step in the present
multi-decomposition technique is generally to calculate the norm
$\nu$.  In sections~\ref{e.examples} and \ref{urn}, we list several
examples where this calculation is feasible exactly. The idea is to
construct a matrix which has the same eigenvalues as $\Pi$, which is
much smaller than $\Pi$ and the elements of which depend only on the
equilibrium distribution $\vect{\pi}$ and the subsets $\Omega_{k;a}$.
This is an important point: $\nu$ is calculated {\em at equilibrium}
and the spectral gap of $P$, that is its {\em dynamical} rate of
convergence, is eventually monitored by this {\em equilibrium}
quantity.

However, even in the cases where this calculation is out of reach, the
present technique is also a great advance as compared to direct
diagonalization of the matrix $P$ to obtain $g(P)$, because in order
to calculate $\nu$ numerically, one has to diagonalize a much smaller
matrix than $P$. This point of view is used at the end of
section~\ref{CTP} to support a conjecture about the ``Contingency
table problem''.

As it was already remarked in section~\ref{defs}, given a
decomposition, $(\Omega_{k;a})_{a=1,\ldots,A_k}$, since the different
subsets have been ``disconnected'', the vector space $E_k$ of
eigenstates corresponding to the eigenvalue 1 is degenerate of
dimension $A_k$. In order to write the projections $\eeq{k} -
\vect{\pi} = \Pi_k (\vect{e}- \vect{\pi})$, which belong to $E_k$, we
need a suitable basis of $E_k$.  For each $k$ and each $a$, we define
the vector $\vect{g}_{k;a}$, the coordinates of which are equal to
$g_{k;a}(x)=\pi(x)$ for any state $x \in \Omega_{k;a}$
and to 0 anywhere else. Note that 
\begin{equation}
\| \vect{g}_{k;a} \| = \left(\sum_{x \in \Omega_{k;a}} \pi(x) \right)^{1/2}.
\end{equation}
We set
\begin{equation}
\vect{u}_{k;a}=\frac{\vect{g}_{k;a}}{ \| \vect{g}_{k;a} \|}.
\end{equation}
Then 
$(\vect{u}_{k;a})_{a=1,\ldots,A_k}$ is an orthonormal basis 
of $E_k$ and by definition of $\Pi_k$ (section~\ref{defs}),
\begin{equation}
\Pi_k (\vect{f}) = \sum_{a=1}^{A_k} \;
\langle \vect{f} | \vect{u}_{k;a} \rangle \; \vect{u}_{k;a}
\end{equation}
for any vector $\vect{f}$. Hence
\begin{equation}
\Pi (\vect{f}) = \sum_k \Pi_k (\vect{f}) = 
\sum_{k=1}^m \sum_{a=1}^{A_k} \;
\langle \vect{f} | \vect{u}_{k;a} \rangle \; \vect{u}_{k;a}
\end{equation}
Let us denote by $U$ the matrix which has the different vectors
$\vect{u}_{k;a}$ for all $k$ and all $a$ as column vectors; this
matrix has $N$ lines and $\sum_k A_k$ columns (the total number of
subsets $\Omega_{k;a}$). We also denote by $\PC$ the diagonal matrix
of diagonal elements the $\pi(x)$, $x\in\Omega$. Then
\begin{equation}
\Pi (\vect{f}) = U U^{t} \PC^{-1} \vect{f}
\end{equation}
where $U^{t}$ is the transpose of $U$ with the vectors
$\vect{u}_{k;a}$ as line vectors. We are interested in the eigenvalues
of $\Pi$ and therefore of $U U^{t} \PC^{-1}$.  Now given any two
matrices $A$ and $B$, a common theorem in basic linear
algebra~\cite{Mehta} says that $AB$ and $BA$ have the same 
non-zero eigenvalues, even if $A$ and $B$ are not square.
As a consequence, $\Pi$ has the same non-zero eigenvalues as
$\bar{\Pi} = U^{t} \PC^{-1} U$, the coefficients of which are
\begin{equation}
\bar{\Pi}_{(k;a),(l;b)} =
\langle \vect{u}_{k,a} | \vect{u}_{l,b} \rangle =
\frac{\sum_{x \in \Omega_{k;a} \cap \Omega_{l;b}} \pi(x)}
{\left( \sum_{x \in \Omega_{k;a}} \pi(x) \right)^{1/2}
\left( \sum_{x \in \Omega_{l;b}} \pi(x) \right)^{1/2}}
\label{pibar}
\end{equation}
This matrix is positive semi-definite; its eigenvalue are non-negative.

\medskip

\begin{theo}
\label{th2}
Let $(\Omega_{1;a_1}),(\Omega_{2;a_2}),\ldots, (\Omega_{m;a_m})$ be a
non-degenerate $(m,r,\nu)$-multi-decomposition of a Markov chain
$\MC$, of stationary distribution $\pi(x)$, the norm of the associated
multi-projection is the second largest eigenvalue of the matrix
$\bar{\Pi}$ defined by eq.~(\ref{pibar}).
\end{theo}

The dimension of the square matrix $\bar{\Pi}$ is $\sum_k A_k$. This
quantity is in general much smaller than the dimension $N$ of $P$. If
the distribution $\vect{\pi}$ is uniform (for example at infinite
temperature), then
\begin{equation}
\bar{\Pi}_{(k;a),(l;b)} = \frac{| \Omega_{k;a} \cap \Omega_{l;b} | }
{|\Omega_{k;a}|^{1/2} |\Omega_{l;b}|^{1/2}},
\label{uniform}
\end{equation}
where $|S|$ stands for the cardinality of any set $S$. One may also
choose to work in a different basis where the above matrix may have a
more practical expression: if $\QC$ is the diagonal matrix with
diagonal coefficients $|\Omega_{k;a}|^{1/2}$, then $\hat{\Pi} = \QC
\bar{\Pi} \QC^{-1}$ has the same eigenvalues as $\bar{\Pi}$ and
$\Pi$. Its coefficients are:
\begin{equation}
\hat{\Pi}_{(k;a),(l;b)} = \frac{| \Omega_{k;a} \cap \Omega_{l;b} | }
{|\Omega_{l;b}|}.
\label{uniform2}
\end{equation}
The same kind of transformation can also be used in the case where $\vect{\pi}$
is not uniform:
\begin{equation}
\hat{\Pi}_{(k;a),(l;b)} =
\frac{\sum_{x \in \Omega_{k;a} \cap \Omega_{l;b}} \pi(x)}
{ \sum_{x \in \Omega_{l;b}} \pi(x)}
\label{pihat}
\end{equation}
This is the conditional probability, at 
equilibrium, that the system is in the subset $\Omega_{k;a}$ given
that it is in the subset $\Omega_{l;b}$.

\section{Elementary examples}
\label{e.examples}

In this section, we display several examples where the
multi-decomposition technique is particularly efficient. In these
three examples, the lower bound provided by the method can be compared
to exact values of the gap calculated by independent means. In these
three cases, the lower bound of theorem~\ref{th1} turns out to be an
equality. 

\subsection{Random walk on a $m$-dimensional cube}
\label{mdimcube}

We consider a random walker on a $m$-dimensional hyper-cube. We first
derive exactly the spectral gap of this Markov chain. Then we prove
that the lower bound provided by the multi-decomposition method is in
fact an equality.

Our random walker moves on the vertices of a cube,
$\Omega_m=\{0,1\}^m$. He moves through the edges of the cube. His
position is denoted by $(x_1,x_2,\ldots,x_m)$. At each step, he
chooses randomly an index $i$ among $m$ as well as a direction $b=\pm
1$. If the move $x_i \rightarrow x_i + b$ is possible then he performs
this move, else he stays at the same place. The allowed transition
rates are all equal to $1/2m$. Then the transition matrix $P^{(m)}$ of
this chain can be written as a tensor product of $m$ one-dimensional
walkers on $\{0,1\}$:
\begin{equation}
P^{(m)} = {1 \over m} \sum_{k=1}^m (\openone \otimes \ldots \otimes
\openone \otimes t_k \otimes \openone \otimes \ldots \otimes
\openone),
\label{tensorprod}
\end{equation}
because the walker can choose one of the $m$ directions with equal
probability $1/m$ and then it performs a one-dimensional walk in that
direction. The transition matrix of each one-dimensional walker 
in the direction $k$ is
\begin{equation}
t_k = t = \frac{1}{2} \left(
\begin{array}{cc}
1 & 1 \\
1 & 1
\end{array}
\right) .
\end{equation}
The eigenvalues of $t$ are 1 and 0, and  $g(P^{(m)}) = 1/m$.

We now use the multi-decomposition technique in order to compare the
obtained lower bound to this exact value: We relate the spectral gap
of $P^{(m)}$ to that of a random walker on a $(m-1)$-dimensional
hyper-cube $\Omega_{m-1}$. There are $m$ natural ways to build the
decompositions, by preventing moves in one (and only one) of the $m$
directions of space. Then there are 2 subsets in each decomposition, 
and they are isomorphic to a hyper-cube $\Omega_{m-1}$.

Furthermore, each transition belongs to exactly $m-1$ decompositions
among $m$, more precisely to every decomposition save the one
where this transition is forbidden. Therefore this multi-decomposition
is regular of redundancy $m-1$. 

Now we consider the multi-projection associated with this
multi-decomposition. Let us prove with the help of
relation~(\ref{uniform}) (and theorem~\ref{th2}) that its norm $\nu
=1$. In this case, $|\Omega_{k;a}|=2^{m-1}$, the cardinality of a
$(m-1)$-dimensional hyper-cube, for all $k=1,\ldots,m$ and all $a=1,2$;
and $| \Omega_{k;a} \cap \Omega_{l;b} | = 2^{m-2}$ whenever $k \neq
l$.  Hence $\bar{\Pi}_{(k;a),(l;b)} = 1/2$ if $k \neq l$ and
$\bar{\Pi}$ is a block-matrix of $2 \times 2$ blocks. Diagonal blocks
are equal to the identity and non-diagonal ones to $t$.  It is an
elementary exercise to compute the eigenvalues of such a matrix: $m$
is the largest eigenvalue, 1 is $m$ times degenerate, and the $m-1$
remaining eigenvalues are equal to 0. We conclude that
$\nu=1$. According to theorem~\ref{th1}, $g(P^{(m)}) \geq
\inf_{k=1,\ldots,m} g(P_k)$. All matrices $P_k$ have the same gap.

Now it is time to draw the reader's attention to the following
important point: the (allowed) non-diagonal transition rates in
matrices $P^{(m)}$ and $P_k$ are equal. Therefore $P_k$ is the direct
sum (see eq. (\ref{sommedir})) of 2 transition matrices of a
$(m-1)$-dimensional random walker, but with non-diagonal transition
rates $1/2m$ instead of $1/2(m-1)$ in $P^{(m-1)}$. As a consequence,
\begin{equation}
g(P_k)=\frac{m-1}{m} \; g( P^{(m-1)})
\end{equation}
and
\begin{equation}
g(P^{(m)}) \geq \frac{m-1}{m} \; g( P^{(m-1)}) \geq \ldots \geq 
\frac{1}{m} \; g( P^{(1)}) = \frac{1}{m}.
\end{equation}
Here, the lower bound calculated {\em via} the multi-decomposition
technique is the exact spectral gap.

\subsection{Random walk on a $m$-dimensional box}
\label{mdimbox}

We extend this calculation to the case considered in introduction,
where the $m$-dimensional box $\Omega_m$ is not necessarily a cube any
longer.  Its side lengths are denoted by $p_1,p_2\ldots,p_m$. We
follow the same line as in the previous example except that now $p_k
\geq 2$.  However, before going on, we need
to clarify the question of boundary conditions, which was temporarily
postponed in the introduction.  In fact, we shall demonstrate the
following striking property: the details of the proof do not depend
precisely on boundary conditions provided the subsets $\Omega_{k;a}$
are the same, as well as the stationary distribution
$\vect{\pi}$. This is the first illustration of a general feature of
the multi-decomposition technique that was already mentioned in
section~\ref{Howto}, and that will be largely discussed in the
following sections: $\nu$ is an equilibrium quantity.

We consider both ``von Neumann'' and periodic boundary conditions.  In
the first case, and as in our previous example, if the walker tries
to move outside the grid, he stays at the same place. In the second
case, the grid lies on a torus and the walker has always $2m$ possible
moves; all the vertices play the same role. In both cases, the allowed
transition rates are equal to $1/4m$. As compared to the previous
examples, where they were set to $1/2m$, these transition rates ensure
that $P(x,x) \geq 1/2$ (In the previous example, this property was due
the the fact that all vertices lay on the boundary).

The exact calculation of the gap of the transition matrix $P^{(m)}$ is
similar to the previous one: $P^{(m)}$ can also be written as a tensor
product of $m$ one-dimensional walkers (eq. (\ref{tensorprod})),
since the walker chooses one of the $m$ directions with equal
probability and performs a one-dimensional walk in that
direction. However, the transition matrix $t_k$ of the one-dimensional
walker now depends both on the side length $p_k$ and the boundary
conditions. For von Neumann ones,
\begin{equation}
t_k = {1 \over 4} \left(
\begin{array}{cccccc}
3 & 1 & 0 & \cdots & \cdots & 0\\
1 & 2 & 1 & 0 & \cdots &  0\\
0 & 1 & 2 & 1 & 0 &  0\\
0 & 0 & \ddots &\ddots &\ddots & 0\\
0 & \cdots &\cdots & 1 & 2 & 1 \\
0 & \cdots &\cdots &\cdots & 1 & 3 \\
\end{array}
\right)
\end{equation}
is a $p_k \times p_k$ matrix. Its eigenvalues are $\lambda_j=\cos(j
\pi /p_k)/2+1/2$, where $j=0,1,\ldots,p_k-1$. The eigenstates of
$P^{(m)}$ are the tensor products of the one-dimensional ones and its
eigenvalues are all the combinations
\begin{equation}
\lambda_{j_1,j_2,\ldots,j_m} = {\lambda_{j_1} + \lambda_{j_2} +
\ldots + \lambda_{j_m} \over m}.
\end{equation}
If $\sup_k p_k=p_{k_0}$, the second largest eigenvalue is obtained
when all the $j_k$'s are set to 0 except for the $k_0$-th one which is set
to 1. Thus $g(P^{(m)})=(1 - \cos(\pi/p_{k_0}))/2m$.  In the case of
periodic boundary conditions, a similar calculation leads to
$g(P_{\mbox{\scriptsize periodic}}^{(m)})=(1 -
\cos(2\pi/p_{k_0}))/2m$, with
\begin{equation}
t_{k,\mbox{\scriptsize periodic}} = {1 \over 4} \left(
\begin{array}{cccccc}
2 & 1 & 0 & \cdots & \cdots & 1\\
1 & 2 & 1 & 0 & \cdots &  0\\
0 & 1 & 2 & 1 & 0 &  0\\
0 & 0 & \ddots &\ddots &\ddots & 0\\
0 & \cdots &\cdots & 1 & 2 & 1 \\
1 & \cdots &\cdots &\cdots & 1 & 2 \\
\end{array}
\right)
\end{equation}
instead of $t_k$.
\begin{figure}[ht]
\begin{center}
\ \psfig{figure=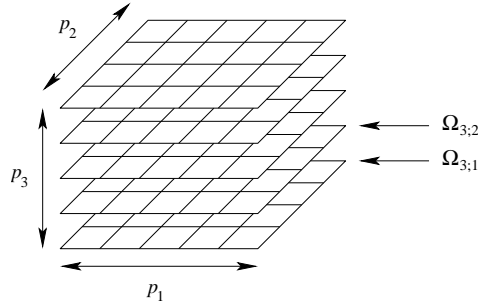,height=4cm} \
\end{center}
\caption{A decomposition (among 3 possible ones) in the case of
a random walker on a three-dimensional cubic grid of
sides $p_1 \times p_2 \times p_3$. The vertical moves, which
are allowed for the three-dimensional walker, are now forbidden 
in this decomposition. The walker is constrained to move 
on one of the $p_3$ two-dimensional grids of sides $p_1 \times p_2$,
denoted by $\Omega_{3;a}$.}
\label{walker3D}
\end{figure}

We now decompose the $m$-dimensional walker in $(m-1)$-dimensional
ones (figure~\ref{walker3D}). The argument is exactly the same as in
the previous section~\ref{mdimcube}, except that $g(P_k)$ now depends
on $k$ since $p_k$ does. For each $k$, the decomposition
$(\Omega_{k,a})$ is obtained by cutting the edges in the direction 
$k$ of the grid. For both boundary conditions, this results in
constraining the walker to move on a $(m-1)$-dimensional grid, with
the same boundary conditions as in the original grid $\Omega_m$.
In addition, the matrix $\bar{\Pi}$ (or $\hat{\Pi}$) is slightly more
complex: we calculate
\begin{equation}
|\Omega_{k;a}|=\prod_{j \neq k} p_j; \ \ \
| \Omega_{k;a} \cap \Omega_{l;b} | = \prod_{j \neq k \atop j \neq l} p_j
\ \ \ \mbox{and} \ \ \ \hat{\Pi}_{(k;a),(l;b)} = \frac{1}{p_k},
\end{equation}
whenever $k \neq l$, whatever $a$ and $b$. The eigenvalues of
$\hat{\Pi}$ can also be exactly calculated and one also gets $\nu=1$.
The redundancy is $r=m-1$ as it is in the previous example.
Theorem~\ref{th1} can be applied and one gets the gap relation
$g(P^{(m)}) \geq \inf_k g(P_k)$.  We recall that $P_k$ is the direct
sum (see eq. (\ref{sommedir})) of $p_k$ transition matrices of a
$(m-1)$-dimensional random walker, with transition rates $1/4m$ instead
of $1/4(m-1)$. Taking this fact into account, one gets by induction
\begin{equation}
g(P^{(m)}) \geq \frac{1}{2m} \; \inf_{k=1,\ldots,m} (1-\cos(\pi/p_k))
\end{equation}
for von Neumann boundary conditions. The exact gap was calculated
above and this lower bound is in fact exact. 

The conclusion is the same for periodic boundary conditions: since the
subsets $\Omega_{k;a}$ are the same as in the von Neumann case, the
matrix $\hat{\Pi}$ is identical for both boundary
conditions. Therefore $\nu$ and $r$ are also equal in both cases, and
the gap relations are identical. Finally, the two proof are exactly
similar except that the first steps of the proof by induction, that is
to say the spectral gaps of the {\em one}-dimensional random walkers,
differ. One finally also gets the exact gap calculated above for
periodic boundary conditions.

\subsection{Random walk on the symmetric group with random transpositions}
\label{Sm}

This Markov chain belongs to the ``Card shuffling
models''~\cite{Aldous81}. In this example, we re-derive the spectral
gap which is usually calculated by group representation theory
arguments \cite{Diaconis81}. In this case, the multi-decomposition
technique is remarkably simple and efficient and provides again an exact lower
bound.

Given an integer $m\geq 2$ and the symmetric group $S_m$, the ``Random
transposition Markov chain''~\cite{Aldous81,Diaconis81} is defined on
$\Omega = S_m$ as follows: given a state (a permutation) $\sigma =
(\sigma_1,\sigma_2,\ldots,\sigma_m)$, pick up uniformly two different
positions $k$ and $l$ at random, then transpose $\sigma_k$ and
$\sigma_l$ with probability 1/2 (so that $P(x,x) \geq 1/2$). This
Markov chain converges towards the uniform distribution on $S_m$
(see~\cite{Aldous81,Diaconis81}). The transition matrix $P^{(m)}$ is
defined as follows: if two different permutations $x$ and $y$ differ
by a single transposition then $P(x,y)=1/[m(m-1)]$. In
reference~\cite{Diaconis81}, it is proven that 
\begin{equation}
g(P^{(m)}) = \frac{1}{m-1} 
\end{equation}
with arguments based on the theory of representations of $S_m$ (the
result is slightly different in reference~\cite{Diaconis81} because
transition rates are slightly different).

Now we decompose the random walk on $S_m$ in random walks on $S_{m-1}$
in order to compute inductively the lower bound on the gap. The
decomposition is natural: for a given $k=1,\ldots,m$, we fix the
position of $\sigma_k$ and we prevent transpositions involving this
index $k$. The subsets $\Omega_{k;a}$ are defined by the fixed value
$\sigma_k =a$.  The resulting chain on each $\Omega_{k;a}$ is the same
as the original one on a set of cardinality $m-1$, with transition
rates $1/[m(m-1)]$ instead of $1/[(m-1)(m-2)]$. Now we calculate
$\bar{\Pi}$:
\begin{equation}
|\Omega_{k;a}|= (m-1)! \ ; \ \ \ |\Omega_{k;a} \cap \Omega_{l;b} | =
(m-2)! \ ; \ \ \ \bar{\Pi}_{(k;a),(l;b)} =
\frac{1}{m-1},
\end{equation}
when $k \neq l$ and $a \neq b$. If $k \neq l$ but $a = b$,
$|\Omega_{k;a} \cap \Omega_{l;b} |=0$ since $\sigma_k$ and $\sigma_l$
cannot be equal, and $\bar{\Pi}_{(k;a),(l;b)}=0$. 

Hence $\bar{\Pi}$ is a block-matrix with $m$ lines and 
$m$ columns of blocks:
\begin{equation}
\bar{\Pi} = \left(
\begin{array}{c|c|c|c|c}
\openone & B & \cdots & \cdots & B \\ \hline
B & \openone & B & \cdots & B \\ \hline
\vdots & \ddots & \ddots &  \ddots & \vdots \\ \hline
B & \cdots & B & \openone & B \\ \hline
B & \cdots & \cdots & B & \openone
\end{array}
\right) ,
\label{Bblock}
\end{equation}
where each block is itself a $m \times m$ matrix and
\begin{equation}
B=\left(
\begin{array}{ccccc}
0 & \frac{1}{m-1} & \cdots & \cdots & \frac{1}{m-1} \\
\frac{1}{m-1} & 0 & \frac{1}{m-1} & \cdots & \frac{1}{m-1} \\
\vdots & \ddots & \ddots &  \ddots & \vdots \\
\frac{1}{m-1} & \cdots & \frac{1}{m-1} & 0 & \frac{1}{m-1} \\
\frac{1}{m-1} & \cdots & \cdots & \frac{1}{m-1} & 0 
\end{array} \right).
\end{equation}
The eigenvalues of $B$ are 1 (non-degenerate) and $-1/(m-1)$, $m-1$
times degenerate, with respective eigenvectors $X_0=(1,\ldots,1)$ and
$X_i=(0,\ldots,0,1,-1,0,\ldots,0)$, $i=1,\ldots,m-1$. Then one
computes the $m^2$ eigenvectors of $\bar{\Pi}$: the $(m-1)^2$ vectors
$(0,\ldots,0,X_i,-X_i,0,\ldots,0)$, where $i>0$, with the same
eigenvalue $m/(m-1)$; the $(m-1)$ vectors
$(0,\ldots,0,X_0,-X_0,0,\ldots,0)$, with eigenvalue $0$; the $(m-1)$
vectors $(X_i,X_i,\ldots,X_i)$ with eigenvalue $0$; and the vector
$(X_0,X_0,\ldots,X_0)$ with eigenvalue $m$. Finally, the
multi-decomposition is non-degenerate and
\begin{equation}
\nu=\frac{m}{m-1}.
\end{equation}
This value tends to 1 when $m$ is large but is not equal to 1. This
nuance is crucial in order to get the good behavior of the gap with
$m$.

On the other hand, this multi-decomposition is regular with redundancy
$r=m-2$. Indeed, each transposition of indices $k$ and $l$ is
forbidden in exactly two decompositions, namely $(\Omega_{k;a})$ and
$(\Omega_{l;b})$. Then theorem~\ref{th1} provides the gap relation:
\begin{equation}
g(P^{(m)}) \geq \frac{m}{m-1} \; \inf_k g(P_k).
\end{equation}
Taking into account the difference of transition rates in $P_k$ and
$P^{(m-1)}$, we have $g(P_k) = (m-2)/m \ g(P^{(m-1)}) $ and 
\begin{equation}
g(P^{(m)}) \geq \frac{m-2}{m-1} \;g(P^{(m-1)}) \geq \ldots \geq
\frac{1}{m-1} \;g(P^{(2)}) = \frac{1}{m-1}
\end{equation}
because $g(P^{(2)})=1$. Once again, this 
bound is the exact gap.

\section{Urn problems}
\label{urn}

In this section, we show that the multi-decomposition technique can be
applied to a sub-class of the larger class of ``Urn
models''~\cite{Urns,Monkey}. The latter have been designed as toy models for
glassy dynamics, to which belongs the maybe more notorious
``Backgammon model''~\cite{Ritort95}. For the first time, we
explicitly introduce temperature in our method. However, 
for the sake of simplicity, we shall essentially focus on
dynamics at vanishing temperature $T \ra 0$.

In urn models, $m$ identical urns contain $n$ identical
(distinguishable or not) particles, or balls. In the historical
Ehrenfest urn model, $m=2$ and $n$ is large. Then balls can be
exchanged from one urn to another, which will define the elementary
moves. The different rules of exchange characterize the different urn
models. In general, they are defined so that the system prefers to
have the maximum of empty urns, and eventually only an urn containing
all the balls. Usually, these urn models are endowed with a
Hamiltonian: the energy $E$ of a configuration is the
number of non-empty boxes.

We shall focus on two different models governed by a Metropolis
algorithm at $T\ra0$ based upon the previous Hamiltonian. These two
models are identical, except that balls are indistinguishable in the
first model and distinguishable in the second one. We shall see that
this difference has dramatic consequences on their dynamics. The
first model is usually known as the ``Monkey model'' and the second
one as the ``Backgammon model''. Their elementary moves are defined as
follows: 
\begin{itemize}
\item[(i)] ``Monkey model''~\cite{Monkey}: balls are {\em
indistinguishable;} two distinct urns are chosen uniformly at random
among the $m$ possible ones, a ``departure'' urn $D$, and an
``arrival'' urn $A$. If a ball can be removed from $D$ to $A$ without
increasing the energy $E$, that is to say if $D$ {\em and} $A$ are not
empty, this move is performed\footnote[1]{In the monkey model of
Ref.~\cite{Monkey}, the arrival box is non-empty; we get rid of
this constraint for the sake of convenience with respect to
our method; this modification does not alter the overall conclusions.};
\item[(ii)] ``Backgammon model''~\cite{Ritort95}: balls are {\em
distinguishable;} among all the balls contained in the $m$ urns, one
ball $b$ is chosen at random and its urn is denoted by $D$. A second urn is
chosen uniformly at random {\em among the $m-1$ remaining ones}.
denoted by $A$. The ball $b$ is taken from $D$ to $A$ if this
move does not increase the energy (that is if $A$ is not empty).
\end{itemize}

By construction, in both models, the number of empty urns cannot but
increase. The dynamics is frozen when all the urns are empty save one.
We can already foresee that model (ii) will be slower to reach this
frozen state than model (i) because it is more difficult to empty an
urn entirely: when one urn is nearly empty, the ball $b$ belongs to
the other urns with much higher probability.

Note that at $T>0$, this Metropolis algorithm must be modified as
follows~\cite{Newman}: an elementary move is accepted with probability
$\min(1,\exp(-\Delta E)/T)$, where $\Delta E$ is the energy variation
is the move {\em was} accepted. The canonical distributions 
corresponding to the above Hamiltonian are:
\begin{equation}
\pi_{\mbox{\scriptsize
(i)}}(n_1,\ldots,n_m)=\frac{1}{Z_{\mbox{\scriptsize (i)}}} \; \exp(-E/T)
\label{distrib.i}
\end{equation}
for model (i), and
\begin{equation}
\pi_{\mbox{\scriptsize
(ii)}}(n_1,\ldots,n_m)=\frac{1}{Z_{\mbox{\scriptsize (ii)}}}
\frac{n!}{n_1!\ldots n_m!}\; \exp(-E/T)
\label{distrib.ii}
\end{equation}
for model (ii), where $n_i$ represents the number of balls in the urn 
number $i$. These distributions satisfy the detailed balance~(\ref{bilan})
for the Metropolis Markov chain.

First we treat the backgammon model which displays
slow dynamics. The monkey model will be studied at the end of the
section. We will prove below that the spectral gap of the backgammon
model at $T \ra 0$ decays exponentially with $n$. By contrast, the
monkey model is rapid: $1/g(P)$ is polynomial in $m$ and $n$. As a
consequence, all the glassy character of the backgammon model comes
from the distinguishable character of its particles.

\subsection{Backgammon model: case $m=2$}

To begin with, we study the case of $m=2$ urns, denoted by $U_1$ and
$U_2$. At this stage, let us already discuss the important following
point: at zero temperature, the fundamental state is twice degenerate:
all the balls can be in $U_1$ (state $x_1$) or in $U_2$ (state
$x_2$). In other words, the eigenvalue 1 of $P$ is degenerate
and at vanishingly small temperature, the spectral gap tends to 0 and
the equilibriation time diverges.

But what does this equilibriation time measure? At very low
temperature, the energy landscape is essentially a bistable potential
with an energy barrier of height 1. Thus $1/g(P)$ measures the time
that the system needs to equilibriate between the two fundamental
states $x_1$ and $x_2$, in other words to explore the two potential
minima with almost equal probabilities. Naturally, this time diverges like
$\exp(1/T)$. But we are not interested in such a time but rather in
{\em the typical time to reach one of the two minima}, the
``absorption time'' in Ref.~\cite{Lipowski97}. In other words, at zero
temperature, we consider that the system has reached equilibrium when
one urn is empty (and when all urns are empty save one if there are
several urns). For a finite system, this time remains finite, even at
$T=0$~\cite{Lipowski97,Murthy97}.


Therefore we would like to design a new transition matrix $P'$, the
spectral gap of which measures efficiently the previous time at $T=0$.
A solution consists in mixing up both stationary states as follows:
when the system has reached one of the states $x_1$ or $x_2$ for the
first time at $t_0$, at each time $t>t_0$, he can be flipped to state
$x_1$ or $x_2$ with probability $1/2$. In other words, we set
$P'(x_1,x_2)=P'(x_2,x_1)=P'(x_1,x_1)=P'(x_2,x_2)=1/2$. As a
consequence, at time $t_0+1$, the system can have explored the two
potential minima with equal probabilities. The unique equilibrium
distribution is now $\vect{\pi}=(x_1+x_2)/2$. And the spectral gap
$g(P')$ now measures the equilibriation time we are interested in.
We denote by $P^{(2)}$ the previous matrix $P'$. It is a
$(n+1) \times (n+1)$ matrix, with elements proportional to
the number of balls in the urn $D$:
\begin{equation}
P^{(2)} = \left(
\begin{array}{ccccccc}
\frac{1}{2} & \frac{1}{n} & 0 & \cdots & \cdots & 0 & \frac{1}{2} \\
0 & 0 & \frac{2}{n} & 0 & \cdots & \cdots & 0 \\
0 & \frac{n-1}{n} & \ddots & \frac{3}{n} & 0 & \cdots & 0 \\
\vdots & \ddots & \ddots & \ddots & \ddots & \ddots & \vdots \\
0 & \cdots & 0 & \frac{3}{n} & 0 & \frac{n-1}{n} & 0 \\
0 & \cdots & \cdots & 0 & \frac{2}{n} & 0 & 0 \\
\frac{1}{2} & 0 & \cdots & \cdots & 0 & \frac{1}{n} &\frac{1}{2} 
\end{array}
\right).
\end{equation}
We have not succeeded in diagonalizing this matrix. It is not
specifically our purpose since we principally desire to exemplify how
the multi-decomposition method relates $m$-urn problems to 2-urn
ones. However high precision numerical diagonalization up to $n=200$
shows without ambiguity that $g(P^{(2)}) \simeq 1/2^{n-1}$ when $n$ is
large.  This value seems to be asymptotically exact. We shall see in
the following subsection that all the slow character of the backgammon
model lies in this 2-urn exponentially small gap: adding more urns
will only modify it by a polynomial factor.

\subsection{Backgammon model: case $m>2$}

Now we consider the case $m>2$ and we relate its dynamics to the
previous one $m=2$. We denote the urns by $U_1,\ldots,U_m$.  We apply
the same kind of trick as in the two-urn case to get rid of the
degeneracy of the fundamental state: the $m$ lower energy states, were
all urns are empty save one, are mixed up and form a single
state. Once all urns are empty save one, all the balls can be
transferred to any urn $U_k$ with equal probability $1/m$. This trick
ensures that at $T=0$ the equilibrium state in non-degenerate and that
the spectral gap of the new transition matrix measures the good
equilibriation time, that is to say the typical time needed to empty
all urns save one.

The multi-decomposition is defined as follows: $\Omega_{k,a}$,
$k=1,\ldots,m$, is the set of all configurations where the urn $U_k$
contains exactly $n_k=a$ balls. In this subset, the number of balls of
$U_k$ cannot vary. The integer $a$ can take all the values ranging
from 0 to $n$. Now we prove that the multi-decomposition is not
degenerate and that $\nu=2$, whatever
$m > 2$ and $n \geq 2$. We work with the matrix $\hat{\Pi}$ of
eq.~(\ref{pihat}). 

However, one must keep in mind that all the method relies
on the scalar product~(\ref{scalprod}) where $\pi(x)$ cannot vanish
for any $x$. Therefore we must work at $T>0$ where $\pi(x)>0$ and then
take the limit $T \ra 0$ in the gap relation~(\ref{gaprel}). 
We recall that the coefficients of $\hat{\Pi}$ are the conditional
probabilities at equilibrium that $U_k$ contains $a$ balls
given that $U_l$ contains $b$ balls. When $T \ra 0$, 
in regard to eq. (\ref{distrib.ii}), the system 
condenses on its energy minimum {\em inside} $\Omega_{l;b}$. 
The energy is minimized when all the $n-b$ remaining balls 
are in the same urn. This minimum is $m-1$ times degenerate. Therefore
$a=n-b$ with probability $1/(m-1)$ and $a=0$ with probability $1-1/(m-1)$.

Therefore the limiting matrix $\hat{\Pi}$ is a $m \times m$
block-matrix like (\ref{Bblock}) where each block is now a $(n+1)
\times (n+1)$ matrix and
\begin{equation}
B=\left(
\begin{array}{ccccc}
\frac{m-2}{m-1} & \cdots & \cdots & \frac{m-2}{m-1} & 1 \\ 
0 & \cdots & 0 & \frac{1}{m-1} & 0 \\ 
\vdots & & \psfig{figure=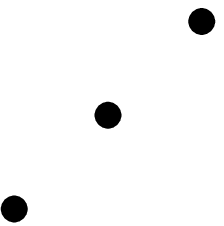,height=3mm} & & \vdots \\
0 & \frac{1}{m-1} & 0 & \cdots & 0 \\
\frac{1}{m-1} & 0 & \cdots & \cdots & 0 
\end{array}
\right) .
\end{equation}
The matrix $B$ can be diagonalized and its eigenvalues are 1
(non-degenerate), $1/(m-1)$ and $-1/(m-1)$. The corresponding
eigenvectors are $(m-1,0,\ldots,0,1)$; the vectors $(1,0,\ldots,0,1)$,
$(0,1,0,\ldots,0,1,0)$, and so forth; the vectors $(1,0,\ldots,0,-1)$,
$(0,1,0,\ldots,0,-1,0)$, and so forth. The exact number of vectors of
the two last species depends on the parity of $n$: if $n$ is even,
one should add the vector $(0,\ldots,0,1,0,\ldots,0)$, where the
one is at the middle of the vector. The corresponding eigenvalue is
$1/(m-1)$. Then the same kind of calculation as in the previous
example \ref{Sm} leads to the conclusion that the multi-decomposition
is non-degenerate and that $\nu=2$.

Like in the previous example, this multi-decomposition is regular with
redundancy $r=m-2$ since each transposition of indices $k$ and $l$ is
forbidden in exactly two decompositions. Since $\nu=2$, theorem~\ref{th1} reads
\begin{equation}
g(P^{(m)}) \geq \inf_{k=1,\ldots,m} \; g(P_k). 
\label{gap_ii0}
\end{equation}
As compared to the previous examples, the situation is more complex
here, because the matrix $P_k$ is {\em not} the direct sum of
identical matrices. According to the number $n_k=a$ of balls that are
stuck in the urn $U_k$, the number of particles that are effectively
involved in the dynamics on the subset $\Omega_{k;a}$ varies.  It is
equal to $n-a$. Now we prove by induction on $m$ that 
\begin{equation}
g(P^{(m)}) \geq \frac{1}{m-1} \;g(P^{(2)}) 
\label{gap_ii}
\end{equation}
where $g(P^{(2)})$ is the gap of the backgammon model on 2 urns
with the same $n$ balls. 

This relation is clear for $m=2$. Suppose it is true for $m-1$.  The
matrix $P_k$ in (\ref{gap_ii0}) is the direct sum of $n+1$ matrices of
backgammon models on $m-1$ urns with $n-a$ balls,
$a=0,\ldots,n$. However, their non-diagonal transition rates must be
re-scaled by a factor $(n-a)/n$ to get $P_k$, so as to take into account
that the {\em number of balls} differs in both matrices; and by a
factor $(m-2)/(m-1)$ so as to take into account the {\em number of
arrival urns}. Moreover, up to urn permutations, all matrices $P_k$
are identical and they have the same gaps. Hence
\begin{eqnarray}
g(P^{(m)}) & \geq & \inf_{a=0,\ldots,n-1} \; 
\frac{n-a}{n} \; \frac{m-2}{m-1} \; g(P_{n-a}^{(m-1)}) \\ \nonumber
 & = & \frac{1}{m-1} \; \inf_{a<n} \; \frac{n-a}{n} \; 
g(P_{n-a}^{(2)})
\end{eqnarray}
where the subscript $n-a$ means that the system contains $n-a$
balls\footnote[1]{We have excluded the value $a=n$ in the relation.
Indeed if $U_k$ contains $n$ balls, the subset $\Omega_{k;n}$ has only
one trivial state and the Markov chain on $\Omega_{k;0}$ is
degenerate. In eq.~(\ref{zorglub}), this subset does not provide any
eigenvector $\vect{f}_j$ and cannot contribute to $\mu_1$. Therefore
it does not appear in the infimum of (\ref{gaprel}).}. Since
$g(P_{n-a}^{(2)}) \simeq 1/2^{n-a-1}$, the previous infimum is reached when
$a=0$.  Hence (\ref{gap_ii}) is verified.

This relation is of crucial importance in the understanding of the
origin of glassy dynamics: all the slow character (the exponential
decay of the gap with $n$) originates from the $m=2$ case; 
adding additional boxes only divides the gap by a polynomial factor
(see the discussion at the end of this subsection, and
Ref.~\cite{Lipowski97}).

We have tested numerically this gap relation on small systems where
the transition matrix can easily be fully diagonalized numerically,
namely for all $m$ and $n$ such that $m+n\leq 14$ as well as all
$n\leq 20$ for $m \leq 4$. In all cases, the previous inequality turns
out to be an equality. We are led to the following:
\begin{conj}
\label{conjurns}
The inequality~(\ref{gap_ii}) is in fact an equality:
\begin{equation}
g(P^{(m)}) = \frac{1}{m-1} \;g(P^{(2)}).
\end{equation}
\end{conj}

\medskip

\noindent We shall return to this conjecture below.

Usually the time unit in discrete time Markov processes is chosen to
depend on the system size in order to be more physical.  In the
present case, we define the time unit so that the time increment is
$\delta t=1/(m-1)$ at each step of the Markov chain, as in
Ref.~\cite{Ritort95,Lipowski97}. Then on average, one tries one move
per urn and per time unit and one expects that typical times will not
depend on the system size at the large size limit; we will see below
that it is precisely what happens. Spectral gaps must also be re-scaled
to take this point into effect: $g'(P)=(m-1) \; g(P)$.
Relation~(\ref{gap_ii}) becomes
\begin{equation}
g'(P^{(m)}) \geq g'(P^{(2)}).
\label{gap_iiB}
\end{equation}
In this section, we have focused on spectral gaps, whereas other
references~\cite{Ritort95,Lipowski97,Murthy97,Urns,Monkey} consider
absorption or ergodic times $\tau'(m,n)$ instead. Let us make the
assumption that $\tau'(m,n) \simeq 1/g'(P^{(m)})$. In the case $m=2$,
we have seen that $g(P^{(2)}) \simeq 1/2^{n-1}$, and it has been
calculated that $\tau'(2,n) \simeq
2^{n-1}$~\cite{Ritort95,Lipowski97,Murthy97}. Therefore the previous
assumption is exact in this case. Under this assumption, 
relation~(\ref{gap_iiB}) becomes $\tau'(2,n) \geq \tau'(m,n)$.

Since the 2-urn dynamics is naturally
faster than the $m$-urn one (see~\cite{Lipowski97}), we also 
naturally assume
$\tau'(2,n) \leq \tau'(m,n)$. As a conclusion,
\begin{equation}
\tau'(2,n) \simeq \tau'(m,n).
\end{equation}
We recover the conclusion of Lipowski \cite{Lipowski97}, which was
based on numerical simulations. Under the previous assumption,
this relation has the same meaning as conjecture~\ref{conjurns}.

\subsection{Monkey model~\cite{Monkey}}
Now we sketch the proof that by contrast, the monkey model is rapid.
This property is already present when $m=2$. Indeed, in that case, the
number of balls in each urn increases or decreases by one unit with
equal probability at each time step. Therefore this model is
equivalent to a random walker with $n+1$ states with two absorbing
states at its extremities. Its spectral gap is
\begin{equation}
g(P^{(2)}) = 1 - \cos(\pi /n) \simeq \frac{\pi^2}{2} \; \frac{1}{n^2}. 
\end{equation}
Taking care of transition rates, which are now equal to $1/[m(m-1)]$,
all the previous proof can be transposed in the present case. In
particular, in regard to eq.~(\ref{distrib.i}), $\hat{\Pi}$ and $\nu$
are unchanged. One finally gets:
\begin{equation}
g(P^{(m)}) \geq \frac{2}{m(m-1)} g(P^{(2)}) \simeq 
\frac{\pi^2}{m(m-1)} \; \frac{1}{n^2}. 
\end{equation}
The Markov chain is rapid.

\section{Contingency table problem}
\label{CTP}

Finally, we present partial results on a difficult problem from
probability theory: the ``Contingency table problem''.  In this case,
we will not be able to calculate exactly the norm $\nu$ because the
matrices $\bar{\Pi}$ or $\hat{\Pi}$ are not as regular as in the
previous examples. However $\nu$ can be computed numerically for
thousands of different examples because these matrices are
sufficiently small. This leads us to the conjecture that
$\nu \leq 2$.  This upper bound is sufficient to prove that the Markov
chains under consideration are rapid (see equations~(\ref{DS}) to
(\ref{tauprime})). Furthermore, as in section~\ref{mdimbox}, the same
argument can be applied to two different Markov chains commonly
associated with this problem, because $\nu$ is an equilibrium
quantity.

Contingency tables are $m \times n$ matrices of non-negative integers
with given positive row and column sums, which arise in statistics as
two-way tables to store data from sample collection
(see~\cite{Diaconis85,Barlow} for instance). To test correlations
between row and column entries, $\chi^2$-tests are applied to such
tables, but they require to sample (almost) uniformly from the set of
contingency tables with given row and column sums. Since no systematic
way is known to perform this sampling, Monte Carlo Markov chains have
been designed to explore randomly these configuration sets.  We study
two such chains.

Consider a positive integer $\Sigma$ and two sequences of positive
integers: the row sums $r_k$, $k=1,\ldots,m$ and the column sums $c_p$,
$p=1,\ldots,n$ such that
\begin{equation}
\sum_{k=1}^m r_k =  \sum_{p=1}^n c_p = \Sigma.
\end{equation}
Denote by $\Omega_{m,n}$ the set of all the $m \times n$ contingency
tables $(z_{kp})$ such that for all $k$ and all $p$
\begin{equation}
\sum_{p=1}^n z_{kp} = r_k \qquad \mbox{and} \qquad 
\sum_{k=1}^m z_{kp} = c_p.
\end{equation}
An example of contingency table is displayed in figure~\ref{exCTP}.

\subsection{Diaconis and Saloff-Coste's chain~\cite{DS95} $\MC_{DS}$:} 
\label{CTP1}

\begin{figure}[ht]
\begin{center}
\ \psfig{figure=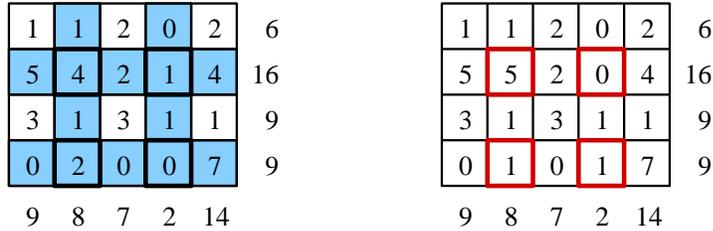,height=3cm} \
\end{center}
\caption{An example of contingency table (left) and of elementary move
as defined in subsection~\protect{\ref{CTP1}}. The row and column sums
are listed on the right and below the table. The two lines and two
columns chosen at random are emphasized, as well as the ``boxes''
$z_{ij}$ that decrease or increase by one unit. This elementary move
does not affect the row and column sums.}
\label{exCTP}
\end{figure}

These authors define a discrete time reversible chain $\MC_{DS}$ on
$\Omega_{m,n}$ as follows: at time $t$ the chain is in the state
$x=(z_{ij})$. With probability 1/2, do nothing (so that $P(x,x) \geq
1/2$); with probability 1/2, pick up uniformly two different rows
$k<l$ as well as two different columns $p<q$ at random. Choose a
number $b \in \{-1,1\}$ uniformly at random. Define $x'=(z'_{ij})$
by $z'_{ij} = z_{ij}$ for all $i$ and $j$ except that:
\begin{eqnarray}
z'_{kp} & = & z_{kp}+b; \\ \nonumber
z'_{kq} & = & z_{kq}-b; \\ \nonumber
z'_{lp} & = & z_{lp}-b; \\ \nonumber
z'_{lq} & = & z_{lq}+b. 
\end{eqnarray}
The alternation of signs ensures that the row and column sums are
conserved (see figure~\ref{exCTP}). If $(z'_{ij})$ is a contingency
table (that is to say if all coefficients are non-negative), then
accept this elementary move; else reject it. This chain $\MC_{DS}$ is
reversible and converges towards the uniform stationary
distribution~\cite{DS95}. The non-zero non-diagonal coefficients of its
transition matrix $P_{DS}^{(m,n)}$ are all equal to
$2/[m(m-1)n(n-1)]$.

The time $\tau$ to reach equilibrium has been calculated in
the particular case of two-row contingency tables~\cite{Hernek98} and
is polynomial in the number of columns $n$ and in the table sum
$\Sigma$.  To extend this result to the general case seems to be a
serious challenge~\cite{Hernek98}.


We build a multi-decomposition by fixing the values of a single row
$k$ (or symmetrically of a single column $p$). Let $\AC_k$ denote the
set of all the possible states in $\Omega$ of the whole row $k$. For
any such state $a \in \AC_k$, $\Omega_{k;a}$ is the set of all
contingency tables in $\Omega_{m,n}$ the $k$-th row of which is
identical to $a$. Then all the subsets $\Omega_{k;a}$ for
$k=1,\ldots,m$ and $a \in \AC_k$ form a suitable multi-decomposition
of $\MC_{DS}$. Indeed, when a row is held fixed, the resulting chain
on $\Omega_{k;a}$ is a chain of type $\MC_{DS}$ on a set of smaller
contingency tables of dimensions $(m-1) \times n$ and of sum
$\Sigma'=\Sigma-r_k$. (In a similar way, should the $p$-th {\em
column} be held fixed, the resulting chain would be a chain of type
$\MC_{DS}$ on a set of smaller contingency tables of dimensions $m
\times (n-1)$ and of sum $\Sigma''=\Sigma-c_p$.) Moreover, this
multi-decomposition is regular of redundancy $r=m-2$ because each move
of row indices $k$ and $l$ is forbidden in exactly two decompositions,
$(\Omega_{k;a})$ and $(\Omega_{l;b})$.

Unfortunately, the calculation of the norm $\nu$ of the multi-projection
$\Pi$ is much more complicated in this case than in the previous examples
because the matrices $\bar{\Pi}$ or $\hat{\Pi}$ lack regularity.
However, thousands of numerical simulations lead us to the following
conjecture:

\medskip

\begin{conj}
\label{conj1}
In the above ``Contingency table problem'', whatever the table
dimensions $m$ and $n$, whatever the table sum $\Sigma$, whatever the
row and column sums $r_k$ and $c_p$, the previous multi-decomposition
is non-degenerate and the following inequality holds: $\nu \leq 2$.
\end{conj}

\medskip

\noindent Note that we already know that conjecture~\ref{conj1} is
satisfied in a particular case: up to transition rates, the Markov
chain on the symmetric group of subsection~\ref{Sm} is 
of $\MC_{DS}$-type when $k=p$ and the $r_k$ and $c_p$ are
all set to 1. Then $\Omega_{m,m}$ is the group of permutation matrices
of size $m$, which is isomorphic to $S_m$. We found $\nu
= m/(m-1) \leq 2$.


If this conjecture is true, theorem~\ref{th1} reads
\begin{equation}
g(P_{DS}^{(m,n)}) \geq \inf_k g(P_k).
\label{gapCT}
\end{equation}
Now $P_k$ is the direct sum (see eq. (\ref{sommedir})) of 
matrices of Markov chains of type $\MC_{DS}$ on
$(m-1)\times n$ contingency tables, except that their transition rates
are equal to $2/[m(m-1)n(n-1)]$ instead of $2/[(m-1)(m-2)n(n-1)]$.
Thus
\begin{equation}
g(P_{DS}^{(m,n)}) \geq \frac{m-2}{m} \; \inf_k g(P^{(m-1,n)}_{k,DS}).
\end{equation}
We keep track of the index $k$ in this relation because the spectral
gap of the restriction of $P$ to the decomposition $(\Omega_{k;a})$
depends on $k$ in general since $r_k$ does. In a similar way, if one
chooses to fix columns indexed by $p$,
\begin{equation}
g(P_{DS}^{(m,n)}) \geq \frac{n-2}{n} \; \inf_p g(P^{(m,n-1)}_{p,DS}).
\end{equation}
Then we understand that inductively, by ``removing'' successively
rows and columns, one will end up with a Markov chain of type
$\MC_{DS}$ on $2 \times 2$ contingency tables and that the following
inequality will hold:
\begin{equation}
g(P_{DS}^{(m,n)}) \geq \frac{2}{m(m-1)} \frac{2}{n(n-1)} \; 
\inf_{k_1<\ldots<k_{m-2} \atop p_1<\ldots<p_{n-2}}
g(P^{(2,2)}_{\{k_i;p_j\} }).
\label{gapDSfin}
\end{equation}
In this inequality, $P^{(2,2)}_{\{k_i;p_j\} }$ is the transition
matrix of the Markov chain of type $\MC_{DS}$ on $2 \times 2$
contingency tables obtained by fixing successively $m-2$ rows and
$n-2$ columns of the initial $m \times n$ tables\footnote[1]{It is
possible that, during the previous process, after fixing a certain
amount of rows and columns, one obtains a problem on $m' \times n'$
tables with one (or even more) row (or column) sum equal to 0. In this
case, it is useless to apply the multi-decomposition technique,
because all the entries of that row (or column) are necessarily equal
to 0. By fixing this column, one obtains only one subset
$\Omega_{k;a}$ and inequality (\ref{gapCT}) is trivial in this case.}.

Suppose these $2 \times 2$ contingency tables have row sums $r$ and
$r'$ and column sums $c$ and $c'$ such that $r+r' = c+c'= Z$. These
tables are entirely characterized by one of their coefficients, say
$z_{11}$. This coefficients varies by $\pm 1$ at each step of the
Markov process. Therefore this chain is a one-dimensional random
walker with non-diagonal transition rates equal to $1/4$ and ``von
Neumann'' boundary conditions; $z_{11}$ has a minimal value $\minindex{z}$
and a maximal one $\maxindex{z}$ which depend on the precise values of $c$,
$c'$, $r$ and $r'$. Nevertheless, one is sure that $\minindex{z} \geq 0$
and $\maxindex{z} \leq Z-1 \leq \Sigma-1$, even if these bounds can be very
loose. Therefore in all cases
\begin{equation}
g(P^{(2,2)}_{\{k_i;p_j\} }) \geq \frac{1}{2} \; (1 - \cos(\frac{\pi}{\Sigma})) 
\simeq \frac{\pi^2}{4}
\; \frac{1}{\Sigma^2}
\end{equation}
and inequality~(\ref{gapDSfin}) reads
\begin{equation}
g(P_{DS}^{(m,n)}) \geq \frac{1}{2}
\; \frac{2}{m(m-1)} \; \frac{2}{n(n-1)} \; (1 -\cos(\frac{\pi}{\Sigma})).
\label{DS}
\end{equation}
As a conclusion, $1/g(P_{DS}^{(m,n)})$ is quadratic in $m$, $n$ and
$\Sigma$, provided conjecture~\ref{conj1} is true.

\subsection{Dyer and Greenhill's chain~\cite{DG00} $\MC_{DG}$:}

As compared to $\MC_{DS}$, in this type of chain $\MC_{DG}$, once the two
rows and the two columns have been picked up at random, the new table
is chosen uniformly among all the tables accessible for any value of
$b \in \Zb$.  In practice, $b$ can take all the integral values
between a minimum value $\minindex{b}$ and a maximum one $\maxindex{b}$,
which depend on the table, the two chosen rows and the chosen columns
(see the discussion about $z_{11}$ in the previous subsection). Apart
from this (important) difference, all the previous development can be
applied to the present chain. In particular, the multi-decomposition
is adapted to $\MC_{DG}$. Once again, one notices that the
multi-decomposition technique does not depend on the details of the
chain, provided the stationary distribution $\vect{\pi}$ and the
subsets $\Omega_{k;a}$ are the same.

However, after removing $m-2$ rows and $n-2$ columns, the ``ultimate''
Markov chain of type $\MC_{DG}$ on $2 \times 2$ tables will differ
from the $\MC_{DS}$-type one. Indeed, at each step, the new $2 \times
2$ table is chosen uniformly at random among all the $N$ accessible
ones. Therefore all the coefficients of the transition matrix
$P_{DG}^{(2,2)}$ associated with this chain are equal to $1/N$ and its
gap is 1 whatever $N$, and inequality~(\ref{gapDSfin}) becomes
\begin{equation}
g(P_{DG}^{(m,n)}) \geq 
\; \frac{2}{m(m-1)} \; \frac{2}{n(n-1)}.
\label{DG}
\end{equation}
Hence $1/g(P_{DG}^{(m,n)})$ is also quadratic in $m$ and $n$ but not
in $\Sigma$, provided conjecture~\ref{conj1} is true. Note that it was
already proven rigorously that this chain is polynomial in $n$ when
$m$ is held fixed~\cite{DG00,CDGJM02}.

\subsection{A remark about mixing times $\tau_{var} (\varepsilon)$:}
We use relation~(\ref{epsvar}) to calculate an upper bound of mixing
times $\tau_{var} (\varepsilon)$ for both Markov chains.  Since
$\vect{\pi}$ is uniform, $1/\ln(\pi(x))$ is constant and equal to $\ln
N_{m,n}$, where $N_{m,n}$ denotes the cardinality of $\Omega_{m,n}$.
Now each integral coefficient $z_{kp}$ of the table is non-negative and is
bounded above by the table sum $\Sigma$. Hence $N_{m,n} \leq
\Sigma^{mn}$ and $\ln N_{m,n} \leq mn \ln \Sigma$. As a consequence,
\begin{eqnarray}
\tau_{var} (\varepsilon) & \leq & \frac{1}{2} \; m(m-1) n(n-1)
\frac{1}{1 - \cos(\pi/\Sigma)} \left[ mn \ln \Sigma + \ln
\left(\frac{1}{\varepsilon}\right) \right] \\ \nonumber & \simeq &
\frac{1}{\pi^2} \; m^3 n^3 \Sigma^2 \ln \Sigma +
\frac{1}{\pi^2} \; m^2 n^2 \Sigma^2 \ln 
\left(\frac{1}{\varepsilon}\right)
\end{eqnarray}
for the Diaconis and Sallof-Coste's Markov chain, and
\begin{equation}
\tau'_{var} (\varepsilon) \leq \frac{1}{4} \; m^2(m-1) n^2(n-1) \ln \Sigma
+ \frac{1}{4} \; m(m-1) n(n-1) \ln \left(\frac{1}{\varepsilon}\right)
\label{tauprime}
\end{equation}
for the Dyer and Greenhill's one. As discussed in
references~\cite{DG00,CDGJM02}, the first chain $\MC_{DS}$ is
polynomial in $m$, $n$ and $\Sigma$, whereas the second one $\MC_{DG}$
is polynomial in $m$ and $n$ but only logarithmic in $\Sigma$ (if
conjecture~\ref{conj1} is true). The present method provides a good
understanding of this phenomenon by naturally relating
the dynamics on $m \times n$ tables to the dynamics on
$2 \times 2$ ones.

To close this section, let us mention the existence of generalizations
in higher dimensions of the previous Markov chains on contingency
tables~\cite{Aoki}. It would be interesting to test whether this
technique can be applied to them.

\section{Conclusion}
\label{discussion}

The multi-decomposition technique is a generalization of the ``Chain
decomposition method'' of D. Randall which can be more easy to
implement in certain cases. Instead of a single decomposition, the
method uses several intricate ones. It relates the spectral gap of a
complex Markov chain to the spectral gaps of Markov chain on smaller
subsets of the configuration space. In particular, a key quantity
intervenes in this relation, which is calculated {\em at equilibrium},
whereas the spectral gap is a dynamical quantity.

We have illustrated the potentialities of the multi-decomposition
technique on several examples. The first examples (random walkers and
``card shuffling'') are elementary, in that sense that their spectral
gaps can be calculated by other means. However, they are a good
illustration of the method efficiency.  We have also
exemplified in these examples that the same multi-decomposition can be
applied to different dynamics on the same system.

The last examples are more interesting since they deal with complex
Markov chains of theoretical interest.  The ``Backgammon model'' and
the ``Monkey model'' belong to the wider class of ``Urn models'' and
are related to the physics of glassy systems. We have recovered in a
rigorous way that the former is slow whereas the latter is rapid. This
fundamental difference is only due to the fact that particles are
distinguishable in the first model and indistinguishable in the second
one. These two examples have also been the occasion to introduce
temperature in the method. The ``Contingency table problem'' is
a notoriously difficult problem from probability theory. We have
not been able to solve completely this problem. However, the
computational advantage of the method is illuminating in this 
case: the matrix $\hat{\Pi}$ is much smaller than the original matrix
$P$ and can be numerically diagonalized for much larger systems.
These numerical calculations support a conjecture, the veracity
of which would imply that the Markov chains under interest are rapid.
In this case also, the same multi-decomposition can be
applied to two different dynamics.

In addition, we have remarked that for the four first examples
examined in this paper, the gap relation~(\ref{gaprel}) of
theorem~\ref{th1} turned out to be an equality, since we obtained the
exact gaps at the end of the calculations. It is legitimate to wonder
whether it is always true. Indeed, even if the present version of the
theorem is sufficient to establish that some Markov chains are rapidly
mixing, this would be a stronger result than theorem~\ref{th1}. To
answer this question, we mention that we know an example on a 20-state
Markov chain where (\ref{gaprel}) is only an inequality and not an
equality. Hence the question becomes: is it possible to establish a
criterion to decide whether relation (\ref{gaprel}) is an equality or
not? It is difficult to tackle this question in the frame of our
calculations, because inequalities appear in the demonstration of
theorem~\ref{th1}.


\section*{Acknowledgments}

I wish to express my gratitude to Dana Randall who pointed out to me
the fact that beyond random tilings, the present method might be of
interest in the ``Contingency Table Problem''. I am also grateful to
Vianney Desoutter for constructive discussions about urn models.

 \end{document}